\documentclass[journal,twoside,web]{ieeecolor2}
\usepackage{generic}
\usepackage{cite}
\usepackage{amsmath,amssymb,amsfonts}
\usepackage{algorithmic}
\usepackage{graphicx}
\usepackage{textcomp}
\usepackage{url} 
\usepackage{multirow}
\usepackage{booktabs}
\usepackage{upgreek}
\usepackage{tablefootnote}
\usepackage{float}
\usepackage[switch]{lineno} 
\def\BibTeX{{\rm B\kern-.05em{\sc i\kern-.025em b}\kern-.08em
    T\kern-.1667em\lower.7ex\hbox{E}\kern-.125emX}}
\begin{document}
\title{Freqformer: Frequency-Domain Transformer for 3-D Reconstruction and Quantification of Human Retinal Vasculature}
\author{Lingyun Wang, Bingjie Wang, Jay Chhablani, Jose Alain Sahel, and Shaohua Pi
\thanks{© 2025 IEEE. Personal use of this material is permitted. Permission from IEEE must be obtained for all other uses, in any current or future media, 
including reprinting republishing this material for advertising or promotional purposes, creating new collective works, for resale or redistribution to 
servers or lists, or reuse of any copyrighted component of this work in other works. DOI: 10.1109/TBME.2025.3612332}
\thanks{This work was supported in part by Eye \& Ear Foundation of Pittsburgh, NIH CORE Grant P30 EY08098 and an unrestricted grant from Research to Prevent Blindness to the Department of Ophthalmology, ARPA-H 1AY2AX000056, as well as the University of Pittsburgh Center for Research Computing through the resources provided. Specifically, this work used the H2P cluster, which is supported by NSF award number OAC-2117681. (\textit{Corresponding author: Shaohua Pi.})}
\thanks{Lingyun Wang and Shaohua Pi are with the Department of Ophthalmology, University of Pittsburgh, Pittsburgh, PA 15213, USA, and also with the Department of Bioengineering, University of Pittsburgh, Pittsburgh, PA 15261, USA (e-mail: LingyunWang@pitt.edu; shaohua@pitt.edu).}
\thanks{Bingjie Wang, Jay Chhablani, and Jose Alain Sahel are with the Department of Ophthalmology, University of Pittsburgh, Pittsburgh, PA 15213, USA (e-mail: biw35@pitt.edu; chhablanijk2@upmc.edu; sahelja@upmc.edu).}}

\maketitle

\begin{abstract}
Objective: To achieve accurate 3-D reconstruction and quantitative analysis of human retinal vasculature from a single optical coherence tomography angiography (OCTA) scan. Methods: We introduce Freqformer, a novel Transformer-based model featuring a dual-branch architecture that integrates a Transformer layer for capturing global spatial context with a complex-valued frequency-domain module designed for adaptive frequency enhancement. Freqformer was trained using single depth-plane OCTA images, utilizing volumetrically merged OCTA data as the ground truth. Performance was evaluated quantitatively through 2-D and 3-D image quality metrics. 2-D networks and their 3-D counterparts were compared to assess the differences between enhancing volume slice by slice and enhancing it by 3-D patches. Furthermore, 3-D quantitative vascular metrics were conducted to quantify human retinal vasculature.  Results: Freqformer substantially outperformed existing convolutional neural networks and Transformer-based methods, achieving superior image metrics. Importantly, the enhanced OCTA volumes show strong correlation with the merged volumes on vascular segment count, density, length, and flow index, further underscoring its reliability for quantitative vascular analysis.  3-D counterparts did not yield additional gains in image metrics or downstream 3-D vascular quantification but incurred nearly an order-of-magnitude longer inference time, supporting our 2-D slice-wise enhancement strategy. Additionally, Freqformer showed excellent generalization capability on larger field-of-view scans, surpassing the quality of conventional volumetric merging methods. Conclusion: Freqformer reliably generates high-definition 3-D retinal microvasculature from single-scan OCTA data, enabling precise vascular quantification comparable to standard volumetric merging methods. Significance: Freqformer has promising clinical applications in the early detection, detailed microvasculature analysis, and monitoring of retinal vascular diseases through advanced 3-D quantitative assessments.
\end{abstract}


\section{Introduction}
\IEEEPARstart{R}{etinal} microvasculature is essential for supporting tissue hemostasis and visual function~\cite{taylor2024role}, and of vital importance in signifying retinal vascular diseases at early stage. Recently, optical coherence tomography angiography (OCTA) is emerging in diagnosing retinal diseases~\cite{hwang2015optical,taylor2024role}. However, the current analysis relies mainly on 2-D \textit{en face} images projected from specific slabs, which may oversimplify and limit the complete elucidation of the complex 3-D organization of the retinal microvasculature. The technical challenge in performing direct 3-D analysis of OCTA volumes arises mainly from their limited signal-to-noise ratio (SNR), and therefore poor vascular connectivity.

\begin{figure*}[!t]
\centerline{\includegraphics[width=0.7\textwidth]{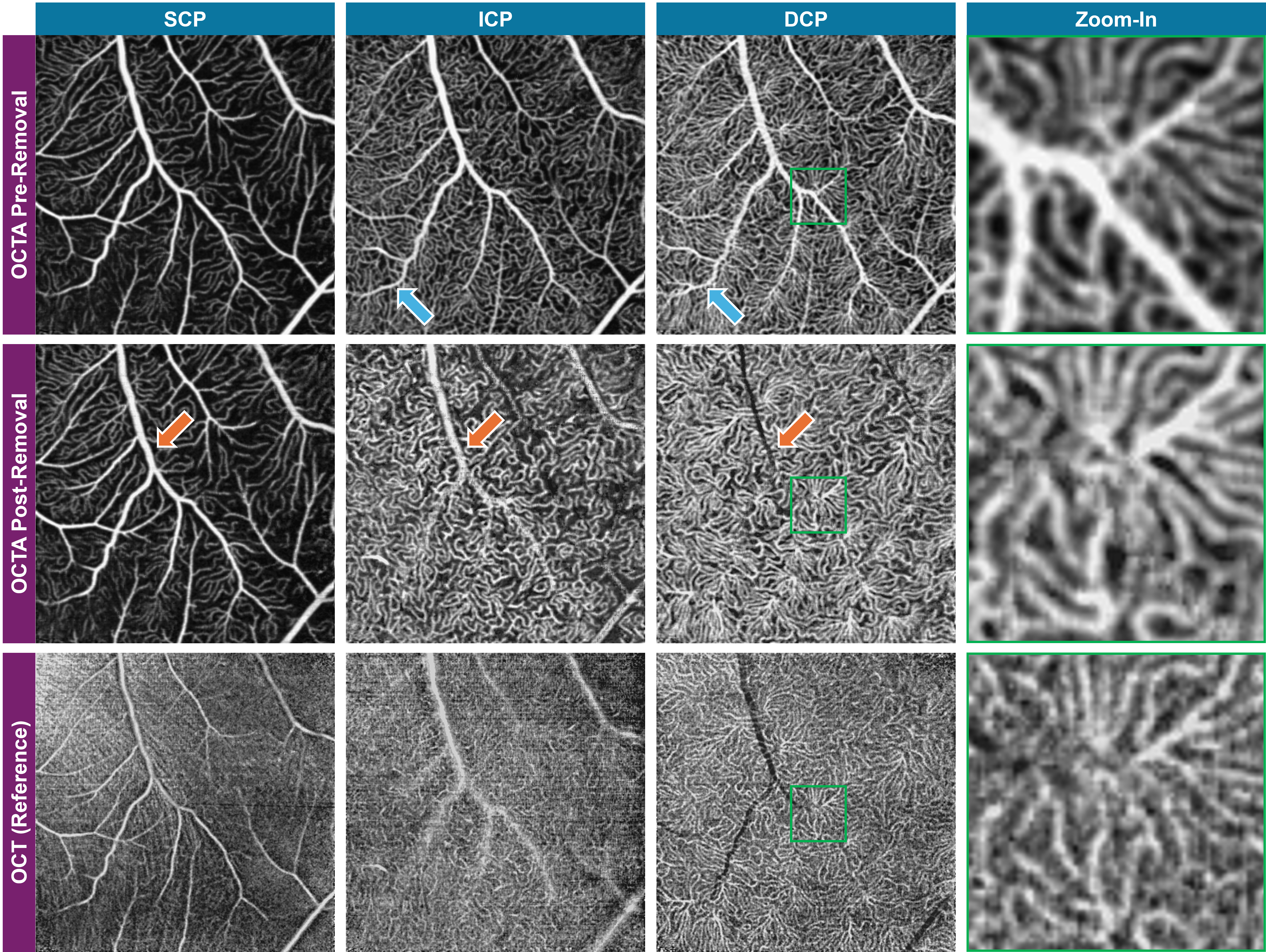}}
\caption{Tail artifact removal of human retinal vasculature in OCTA, with 10-volume merged OCT images as reference to validate the results. Green boxes indicate the region of zoom-in view. Blue arrows indicate the tail artifact as there is no blood vessel in the corresponding OCT image. Orange arrows indicate the shadow artifacts of a large vessel. SCP, superficial capillary plexus. ICP, intermediate capillary plexus. DCP, deep capillary plexus.}
\label{fig1} 
\end{figure*}

Traditional OCTA vascular enhancement is based primarily on shape-based filtering~\cite{meleppat2019multiscale}, wavelet transform~\cite{liu2020using}, and compressive sensing~\cite{wang2022compressive}, which all struggle to preserve fine vascular structures in noisy images. More recently, deep learning-based methods have emerged as powerful tools for OCTA enhancement while limited to 2-D \textit{en face} projection images, and mostly use CNNs. Some models use paired low-resolution (such as 6\(\times\)6 \(\text m \text m^{\text 2}\)) and high-resolution (3\(\times\)3 \(\text m \text m^{\text 2}\)) images to enhance vascular details~\cite{gao2020reconstruction}, while others take a single-scan image as input and use a merged image as ground truth to enhance the angiogram~\cite{kadomoto2020enhanced,gao2025nonperfused}. To the best of our knowledge, the only reported study on 3-D reconstruction of OCTA was performed by training the network on 2-D projection images and then applying it to segment each slice of 3-D OCTA volumes, which demonstrated improved blood vessel visibility~\cite{li2024visualization}. However, its limitations are obvious: a) Time-consuming manual annotation. Although 2-D projection–trained networks can be applied slice-by-slice for 3-D volumes, they often hallucinate vessels or miss faint capillaries when generalizing to slices, which is an artifact of deep learning’s inherent tendency to overfit the statistical priors of their training domain and falter under distribution shift. The resulting masks must therefore require extensive manual refinement, which is still labor-intensive and error-prone. b) Loss of angiogram de-correlation information. Binarized segmentation methods identify vessel structures but discard flow contrast, which is essential for accurate vascular characterization~\cite{su2016calibration}.

We introduce Freqformer as the first deep learning-based approach for 3-D enhancement of OCTA volume, leveraging Transformer-based architectures and frequency-domain learning to overcome the challenges of low SNR and poor vessel connectivity. This pioneering effort opens new avenues for high-resolution volumetric vascular imaging in both clinical and research settings, benefiting the downstream vascular quantification. The key contributions are as follows:

(1) Introduction of Complex-Valued Frequency-Domain Module (CFDM). We propose CFDM, an innovative frequency-domain processing module that selectively enhances critical frequency components, especially mid- and high-frequency components which Transformer overlooks. CFDM uses complex-valued convolutions to compensate for the low-pass filtering property of multi-head attention in Transformer layers and adaptively boost whichever frequencies are most beneficial for feature maps at each stage of the network, ensuring a comprehensive representation of retinal microvasculature from both spatial and frequency domains. This module significantly improves capillary continuity and vessel integrity in depth-plane images.

(2) First Demonstration of Reliable 3-D Capillary Segment-based Analysis in Human Retinas. Freqformer enables direct volumetric reconstruction of retinal vasculature from a single scan, achieving high-fidelity 3-D OCTA enhancement with high consistency with merged volumes. By preserving full 3-D vascular integrity, Freqformer enables quantitative analysis of capillary segments, including vessel count, density, length and flow index.

\section{Method}
\subsection{Data Acquisition and Preprocessing}
This study involves healthy human subjects recruited from UPMC Vision Institute, Department of Ophthalmology at the University of Pittsburgh, Pittsburgh, PA USA. We enrolled 24  adults (13 male/11 female), mean age 30.5 ± 5.8 years (range 22–45). Inclusion criteria were clear ocular media, reliable fixation, and no history of ocular disease. Exclusion criteria were ZEISS OCT signal strength $<$8/10, severe motion artifacts, or failure of 10-volume registration for generating ground truth. The study was approved by the institutional review board and all participants provided their written informed consent. The study was conducted following strict ethical guidelines, with all participants signing informed consent forms, in compliance with the Declaration of Helsinki. Ethical approval was obtained from the institutional review board of the University of Pittsburgh. 3\(\times\)3 \(\text{mm}^{\text 2} \) OCT/A scans are acquired across various retinal regions using a PLEX® Elite 9000 swept-source OCT (ZEISS, Dublin, CA) at a scanning rate of 100 kHz. Each imaging session generated 400 A-lines per B-scan, and 400 B-scan per volume, with a total of 10 repeated volumes.

To generate ground truth, the repeated volumes from each session are registered and merged using recently developed algorithms~\cite{wang2023volumetrically}. Briefly, in the lateral direction, non-rigid registration is performed by aligning retinal vascular patterns in \textit{en face} images. In the axial direction, a depth transformation matrix is computed by calculating the cross-correlation of A-line profiles and identifying the axial shifts with maximal correlation values. 

\begin{figure*}[!t]
\centerline{\includegraphics[width=0.9\textwidth]{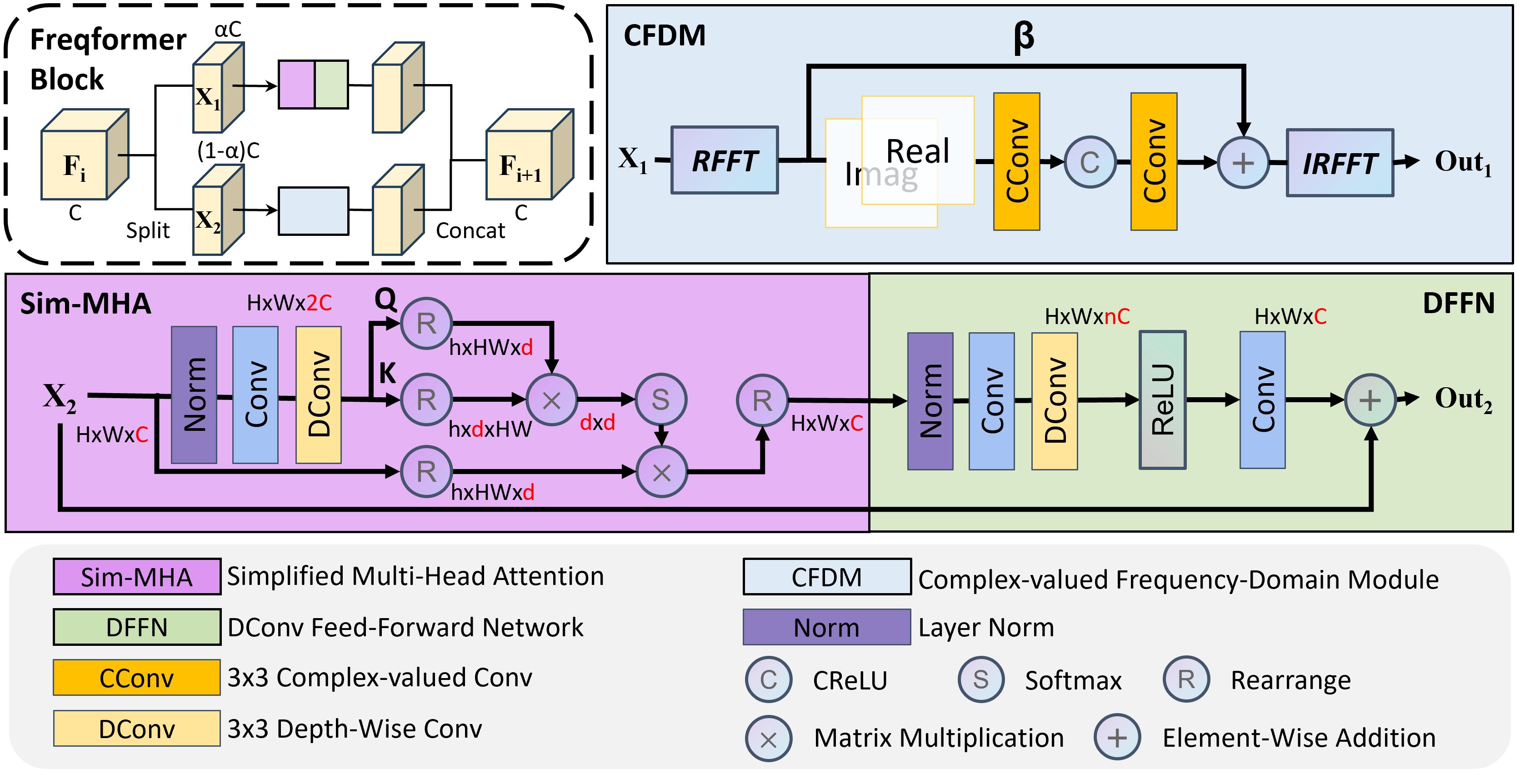}}
\caption{Framework of proposed Freqformer, highlighting the CFDM (light blue color), Sim-MHA (magenta color), and DFFN (light green color) modules within each Freqformer block.}
\label{fig2} 
\end{figure*}

To accurately capture true retinal vasculature in 3-D, removing tail artifacts of blood vessels is required~\cite{hormel2021artifacts}. The tail artifacts are elongated axial streaks that appear immediately posterior to the true blood vessels (as denoted by the blue arrows in Fig.~\ref{fig1}), obscuring true vascular structures and leading to inaccurate interpretation of vessel morphology or density. Current methods~\cite{zhang2016projection, liu2019projection, li2022blood} are developed to maximize integrity of capillary plexuses in \textit{en face} images, often neglecting the actual caliber of retinal vessels. As a result, they tend to lose penetrating and connecting vessels/capillaries, limiting their utility for appreciating 3-D retinal vasculature. To address this challenge, we propose a moving average subtraction filter (MASF) method:

\begin{equation}
\begin{aligned}
\text{MASF}(A_{k}) = \sqrt{\mathrm{max}(A_{k}^{2} - \gamma\frac{1}{w} \sum_{i=k-w}^{k-1} A_{i}^{2}, 0)}
\end{aligned}
\label{eq1}
\end{equation}
where \(A_{k}\) denotes the value of the \(k\)-th pixel in an A-line, and \(\gamma\) and \(w\) represent the weight factor and window size, respectively. In this work, \(\gamma\) and \(w\) are empirically set to 0.8 and 11, respectively, as this configuration effectively suppresses tail artifacts while preserving capillary structures. This method leverages the fact that artifact voxels typically have smaller angiogram values than the real vessel voxels located anteriorly due to their weaker de-correlation between the repeated scans, a widely recognized characteristic in OCTA imaging~\cite{zhang2016projection, choi2020mean}. As a result, tail artifacts are effectively reduced, while the capillaries previously buried in the artifacts stand out. As shown in Fig.~\ref{fig1}, the blood vessel highlighted by the orange arrow is faithfully retained, while regions that show no vessel in the structural OCT reference contain no residual flow signal. The result demonstrates that our method avoids both over- and under-suppression of tail artifacts, achieving high fidelity to the true vascular anatomy. It should be noted that in shadow artifact regions (dark streaks beneath true blood vessels)~\cite{golzan2011minimising}, MASF is not able to retrieve the deep capillaries due to strong light-absorption from anterior large blood vessels, consistent with their appearance in OCT images.

\subsection{Freqformer}
\subsubsection{Overall Architecture}
Freqformer employs a 4-stage symmetric encoder-decoder backbone architecture to process single-scan OCTA images (Fig.~\ref{fig2}). Given an input OCTA single depth-plane image \(\mathbf{I} \in \mathbb{R} ^{H\times W\times 1} \), Freqformer first applies a 3\(\times\)3 convolutional layer to extract shallow feature \(\mathbf{F}_{0} \in \mathbb{R} ^{H\times W\times C} \). After that, the encoder-decoder structure is used to extract the deep features. Encoders reduce the spatial resolution of the feature map while doubling the number of channels using pixel-unshuffle operation as the stage grows. Decoders perform the opposite operations to restore the original resolution. At stages 1-3, up-sampled feature maps are concatenated with their corresponding encoder output, and the channel dimension is reduced by half using a 1\(\times\)1 convolutional layer. These skip connections preserve crucial spatial information by minimizing fine detail loss during the down-sampling and up-sampling processes. Additionally, to enhance generalization and regularize the network, randomly path dropping is utilized during training with a drop rate of 0.1. After passing through the encoder-decoder and a 3\(\times\)3 convolutional layer, feature maps undergo further refinement, resulting in \(\mathbf{I}_{r} \in \mathbb{R} ^{H\times W\times 1} \). The enhanced OCTA image is then generated by adding \(\mathbf{I}\) to \(\mathbf{I}_{r}\), effectively merging the original image with the refined deep features to ensure a high-quality output.

Each stage of the encoder-decoder architecture and refinement comprises multiple Freqformer blocks. In each block, the feature map \(\mathbf{F}_{i} \in \mathbb{R} ^{H\times W\times C} \) is split along the channel dimension into two parts \(\mathbf{X}_{1} \in \mathbb{R} ^{H\times W\times \alpha C} \) and \(\mathbf{X}_{2} \in \mathbb{R} ^{H\times W\times (1 - \alpha) C} \), where \(\alpha \) is a hyper‑parameter that determines the channel‑split ratio between the two parallel branches. \(\mathbf{X}_{1}\) is processed by the Transformer layer, which includes the Simplified Multi-Head Attention (Sim-MHA) and Depth-Wise Convolutional Feed-Forward Network (DFFN), to extract global spatial features. \(\mathbf{X}_{2}\) is processed by the Complex-valued Frequency-Domain Module (CFDM) to capture features in the frequency domain. This dual-branch strategy enables Freqformer to leverage complementary spatial and frequency domain features from OCTA images.

\subsubsection{Efficient Transformer Layer}
With feature map \(\mathbf{X} \in \mathbb{R} ^{H\times W\times C} \), conventional multi-head attention (MHA) computes a weighted sum of the values \(\mathbf{\hat{V}} \in \mathbb{R} ^{h\times HW\times d} \), with attention weights \(\mathbf{A} \in \mathbb{R} ^{h\times HW\times HW} \) derived from the dot-product of queries \(\mathbf{\hat{Q}} \in \mathbb{R} ^{h\times HW\times d} \) and transposed keys \(\mathbf{\hat{K}} ^{\top} \in \mathbb{R} ^{h\times d\times HW} \), where \(d\) and \(h\) denote the number of dimensions and heads, respectively. However, MHA's quadratic space complexity \(\mathcal{O} (hH^{2} W^{2} )\) poses significant challenges over the linear complexity of CNNs \(\mathcal{O} (HWC_{out})\) due to the large size of OCTA images. To address this, we introduce cross-covariance (CC) attention~\cite{ali2021xcit}, where attention weights \(\mathbf{A} \in \mathbb{R} ^{h\times C\times C} \) are computed across channels instead of tokens, using the cross-covariance matrix \(\mathbf{\hat{K}}^{\top} \mathbf{\hat{Q}}\). This approach reduces space complexity to \(\mathcal{O} (C^{2} / h)\), scaling linearly with image size.

Additionally, in standard MHA, there are four linear transformations (\(\mathbf{W}_{q}\), \(\mathbf{W}_{k}\), \(\mathbf{W}_{v}\), and \(\mathbf{W}_{projection}\)) which are applied to compute queries (\(\mathbf{Q}\)), keys (\(\mathbf{K}\)), values (\(\mathbf{V}\)), and output features. Here, \(\mathbf{W}_{v}\) and \(\mathbf{W}_{projection}\) are removed as their identity components tend to become dominant as Transformer converges~\cite{trockman2023mimetic}. To improve local context representation, a 3\(\times\)3 depth-wise convolutional layer (DConv) is integrated after the \(\mathbf{W}_{q}\) and \(\mathbf{W}_{k}\) linear transformations, implemented via 1\(\times\)1 convolution. Unlike conventional vanilla convolutional layers, DConv operates in a channel-wise manner, rather than across all channels, thereby reducing computational cost to \(\sim1/2C\). Our Sim-MHA operation is defined as:
\begin{equation}
\begin{aligned}
& \mathbf{X}_{Sim-MHA}= \mathbf{\hat{X}} \cdot \mathrm{softmax} (\mathbf{\hat{K}}^{\top} \mathbf{\hat{Q}}/\tau), \\ 
&\mathbf{K} = \mathrm{LN}(\mathbf{X})\mathbf{W}_{k}  \mathbf{W}_{D} , \\ 
&\mathbf{Q} = \mathrm{LN}(\mathbf{X})\mathbf{W}_{q} \mathbf{W}_{D},
\end{aligned}
\label{eq2}
\end{equation}
where \(\mathbf{W}_{D}\) and \(\mathrm{LN} (\cdot) \) represent the depth-wise convolutional transformation layer and the layer normalization, respectively. \(\tau\) is a trainable scaling factor to prevent gradients vanishing caused by the softmax function. \(\mathbf{\hat{X}}\), \(\mathbf{\hat{K}}\), and \(\mathbf{\hat{Q}}\) are the reshaped form of \(\mathbf{X}\), \(\mathbf{K}\), and \(\mathbf{Q}\) with corresponding dimensions required by the self-attention mechanism~\cite{vaswani2017attention}. 

The feed-forward network (FFN)~\cite{vaswani2017attention} introduces non-linearity to enhance the complex patterns extracted by MHA. Here, by adopting depth-wise convolutional FFN, two linear transformations are applied using 1\(\times\)1 convolutions to expand and reduce channels. A 3\(\times\)3 DConv layer, similar to that in Sim-MHA, is integrated to enhance local context representation. Furthermore, its output is directly connected to the input of Sim-MHA by removing the skip connection within conventional MHA.

\subsubsection{Complex-valued Frequency-Domain Module (CFDM)}
{As mentioned above, the essence of MHA is a weighted sum of values. This summation process, whether simplified or not, tends to diminish high-frequency features~\cite{park2022vision} which are essential in reconstructing the capillaries in OCTA. To address this, we introduce CFDM, which directly learns complex features in the frequency domain~\cite{li2023frequency}.}

\begin{figure*}[!t]
\centerline{\includegraphics[width=1\textwidth]{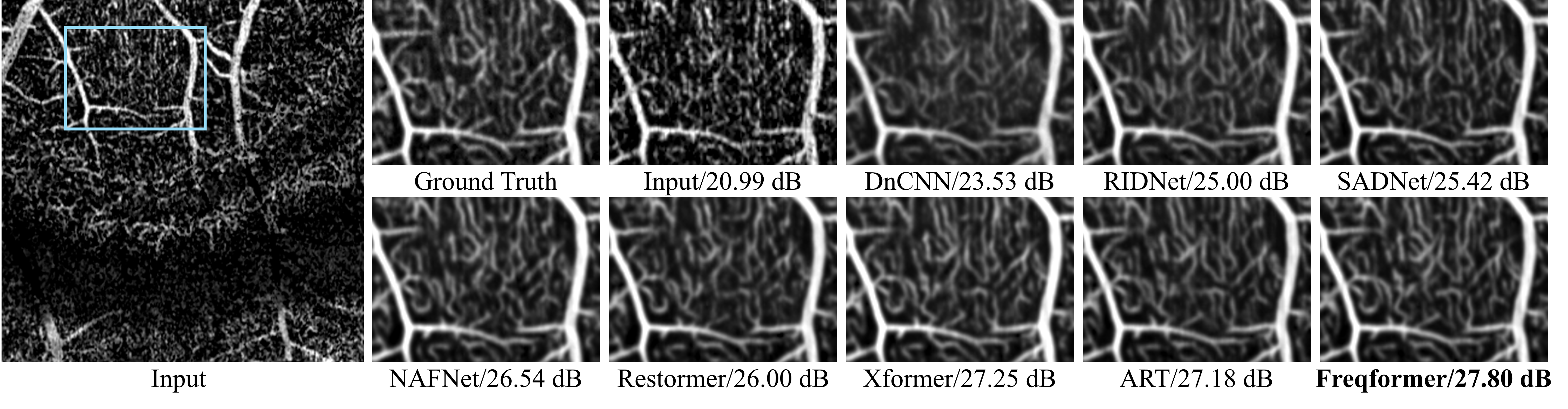}}
\caption{Comparisons of different deep learning enhancement methods with PSNR values listed.}
\label{fig3} 
\end{figure*}

In CFDM, a real-valued fast Fourier transform (RFFT) is applied to the feature map \(\mathbf{X} \in \mathbb{R} ^{H\times W\times C} \), generating a frequency representation \(\mathcal{F}  (\mathbf{X}) \in \mathbb{C} ^{H\times [\frac{W}{2} +1] \times C} \), reducing redundancy and computational cost without losing information due to Hermitian symmetry. After the RFFT, two complex-valued convolutional layers (CConv) are employed to process frequency-domain features~\cite{wang2023axial}. For an input \(\mathcal{F}(\mathbf{X})=\mathbf{x}+i\mathbf{y}\) and a convolution \(\mathbf{W}=\mathbf{u}+i\mathbf{v}\), the operation is \(\mathcal{F}(\mathbf{X})*\mathbf{W}=(\mathbf{x}*\mathbf{u}-\mathbf{y}*\mathbf{v})+i(\mathbf{x}*\mathbf{v}+\mathbf{y}*\mathbf{u})\). Therefore, the gradient of back-propagation can be expressed as:
\begin{equation}
\frac{\partial L(\mathcal{F}(\mathbf{X})*\mathbf{W})}{\partial \mathbf{W}} = \frac{\partial L(\mathcal{F}(\mathbf{X})*\mathbf{W})}{\partial \mathcal{F}(\mathbf{X})*\mathbf{W}}* \mathcal{F}_{flip}(\mathbf{X}),
\label{eq3}
\end{equation}
where \(L(\cdot)\) denotes the loss function and \((\cdot)_{flip}\) represents a $180^{\circ}$ flip of its argument for all spatial axes. The real part of \(\frac{\partial L(\mathcal{F}(\mathbf{X})*\mathbf{W})}{\partial \mathbf{W}}\) is \(\mathfrak{R}\left(\frac{\partial L(\mathcal{F}(\mathbf{X})*\mathbf{W})}{\partial \mathcal{F}(\mathbf{X})*\mathbf{W}}\right) * \mathbf{x}_{flip}-\mathfrak{I}\left(\frac{\partial L(\mathcal{F}(\mathbf{X})*\mathbf{W})}{\partial \mathcal{F}(\mathbf{X})*\mathbf{W}}\right) * \mathbf{y}_{flip}\) while the imaginary part is \(\mathfrak{R}\left(\frac{\partial L(\mathcal{F}(\mathbf{X})*\mathbf{W})}{\partial \mathcal{F}(\mathbf{X})*\mathbf{W}}\right) * \mathbf{y}_{flip}+\mathfrak{I}\left(\frac{\partial L(\mathcal{F}(\mathbf{X})*\mathbf{W})}{\partial \mathcal{F}(\mathbf{X})*\mathbf{W}}\right) * \mathbf{x}_{flip}\). It is worth noting that, although the receptive field of CConv is limited, it still captures global context as each frequency component represents a global characteristic of the entire image. 

Next, CReLU, which performs conventional ReLU on the real part and the imaginary part of feature maps separately, conducts non-linear activation~\cite{trabelsi2017deep}. However, this activation approach can alter the amplitude and phase simultaneously, potentially disrupting the original frequency information. To mitigate this, we introduce a dynamic skip connection between the original complex-valued matrix and the output in the frequency domain, controlled by a trainable parameter \(\beta\). Finally, inverse RFFT is employed to convert the frequency-domain enhanced feature back to spatial domain. 

\subsection{3-D Visualization and Analysis of Retinal Microvasculature}
Next, binarization, color-coding, and 3-D rendering are applied to generate 3-D vascular visualization. Although we enhance OCTA volume slice by slice, all quantification remains volumetric.
Enhanced OCTA volumes are binarized using a threshold (50/255) to separate blood vessels from the background. Skeletonization further refines vessel connectivity by removing small, isolated noise fragments. To enable depth-resolved visualization, the axial boundaries of the superficial, intermediate, and deep capillary plexuses are identified by the corresponding structural OCT using a standard retinal layer segmentation algorithm (SCP: GCL–upper IPL; ICP: lower IPL–upper INL; DCP: lower INL–outer OPL). Vessels are color-coded by plexus: SCP (green), ICP (cyan), DCP (red), and inter-plexus capillaries (blended colors). This color-coded rendering is for visualization only and it is intended to facilitate the tracking of inter-plexus connections. The processed 3-D vasculature is rendered in 3D Slicer, a software platform applying volume rendering techniques to visualize vascular architecture interactively. These approaches allow exploration of detailed vascular spatial distribution, capillary organization, and vessel connectivity. 

Using a similar method described in our previous work~\cite{pi2020retinal}, the 3-D binary masks are skeletonized to shrink vascular diameter and highlight connectivity information. Next, the number of neighboring vascular voxels is counted for each voxel using 26-connectivity in 3-D space to identify bifurcation points. These bifurcation points are unique among body points of blood vessels with more than two neighboring vascular voxels and end points with only one neighboring vascular voxel. Therefore, capillary segments can be isolated and extracted automatically. It should be noted that small fragments with lengths equal or less than six voxels are considered as noise and therefore removed from the identification~\cite{gao2020reconstruction}. 3-D quantitative metrics of blood vessels, including capillary segment count (\(N\)), segment density (\(N/\text{mm}^{3}\)), segment length (voxel) and flow index (average value of OCTA signal, a.u.), are calculated within the inner retina volume. Statistical comparisons among single scans, enhanced scans, and merged scans are performed using one-way ANOVA followed by Tukey's HSD test. A p-value smaller than 0.05 is considered significant in this study. To compare the reliability of each method, the normalized root mean squared error (NRMSE) is calculated between the resulted OCTA volume from each method with the ground truth (merged volume) for each metric.

\section{Experiments and Results}
\subsection{Data Preparation and Implementation Details}
Our training dataset consisted of 3,600 single-merged pairs of depth-plane OCTA images (400 px \(\times\) 400 px) derived from 20 scans across 14 subjects. The dataset was split into 90\% for training and 10\% for validation. For independent evaluation, we constructed a 2-D test set using 400 pairs of images from 24 scans across 10 subjects, ensuring that none of these subjects were included in the training set. Additionally, we created a 3-D test set using full retinal volumes from these 24 scans to comprehensively assess the model’s performance in 3-D space.

The number of Freqformer blocks was set to [2, 4, 4, 6, 4, 4, 2] in the stage of encoder-decoder, and to 4 in the refinement module. Following typical settings in other Transformer methods\cite{zamir2022restormer,zhang2023accurate}, the number of heads in each stage of the encoder-decoder was set to [1, 2, 4, 8, 4, 2, 1], with 1 head used in the refinement module. Additionally, the number of channels was 48 for the feature map input into the encoder-decoder. In the Transformer layer, the Sim-MHA employed bias-free linear transformations, while the DFFN used biased linear transformations, with an expansion factor of 2.66. The channel-split ratio \(\alpha \) was empirically set to 0.5.

The model was trained using AdamW optimizer (\(\beta_{1}\) = 0.9, \(\beta_{2}\) = 0.999, weight decay of \(1e^{-4}\)) and \(L_{1}\) loss. The batch size was set to 1, and the model was trained for 480,000 iterations with an initial learning rate \(1e^{-4}\), halved every 20 epochs until convergence. The reference networks were trained following their original implementation details. All experiments were implemented in Python 3.8 with Pytorch 2.2.0 deep learning framework on a single NVIDIA A100 GPU. 

The performance of Freqformer was assessed on the reconstructed depth-plane images using the aforementioned 2-D test set by three 2-D image quality metrics: peak signal-to-noise ratio (PSNR), structural similarity index measure (SSIM), and learned perceptual image patch similarity (LPIPS, using the pre-trained AlexNet model provided in~\cite{zhang2018unreasonable}). Additionally, we employed two 3-D image quality metrics: 3D-SSIM~\cite{wang2004image} and 3D gradient magnitude similarity deviation (3D-GMSD)~\cite{xue2013gradient} on the 3-D test set to evaluate angiographic fidelity and perceptual consistency of entire retinal microvascular network. The equations are detailed below:
\begin{equation}
\text { PSNR }=10 \log _{10}\left(\frac{\max (\mathbf{I})^{2}}{\frac{1}{N} \sum_{i=1}^{N} \left(\mathbf{I}(i)-\mathbf{I}_{GT}(i)\right)^{2}}\right),
\label{eq4}
\end{equation}

\begin{equation}
\text {SSIM}(p)=\frac{\left(2 \mu_{I}(p) \mu_{I_{GT}}(p)+C_{1}\right)\left(2 \sigma_{II_{GT}}(p)+C_{2}\right)}{\left(\mu_{I}^{2}(p)+\mu_{I_{GT}}^{2}(p)+C_{1}\right)\left(\sigma_{I}^{2}(p)+\sigma_{I_{GT}}^{2}(p)+C_{2}\right)} ,
\label{eq5}
\end{equation}

\begin{equation}
\text{LPIPS} = \sum_{l} \omega_{l} \frac{1}{N_{l}} \sum_{i=1}^{N_{l}}\left\| \phi_l (\mathbf{I}(i)) - \phi_l (\mathbf{I}_{GT}(i)) \right\|_2^2 ,
\label{eq6}
\end{equation}

\begin{equation}
\text{GMS}(p) =\frac{2 m_I(p) m_{I_{GT}}(p) + T}{m_I^2(p) + m_{I_{GT}}^2(p) + T},
\label{eq7}
\end{equation}
where \(\mathbf{I}\) and \(\mathbf{I}_{GT}\) are the enhanced image and the ground truth. \(N\) is the number of pixels in the image. The local means $(\mu_I, \mu_{I_{GT}})$, standard deviations $(\sigma_I, \sigma_{I_{GT}})$, and covariance $\sigma_{II_{GT}}$ are Gaussian‑weighted statistics computed over corresponding $11\times11$ patches centered at the pixel (or voxel) $p$. Weighting is performed with an isotropic Gaussian window of standard deviation $\sigma_{\!w}=1.5$ pixels and unit sum. The stabilizing constants are set to $C_1 = (K_1 L)^2$ and $C_2 = (K_2 L)^2$ with $K_1 = 0.01$, $K_2 = 0.03$, and dynamic range $L = 255$. The SSIM score is obtained by averaging all local SSIM values. For 3D-SSIM, the formulation remains identical, except that all local statistics are computed over an $11\times11\times11$ voxel neighborhood weighted by the same isotropic Gaussian kernel, thereby incorporating correlations along the depth axis. For LPIPS, \(\phi_{l}\) and \(\omega_{l}\) denote the \(l\)-th layer and its weight factor of the pre-trained AlexNet. \(N_{l}\) is the number of pixels in the feature maps at the \(l\)-th layer. The 3D-GMSD score is calculated as the standard deviation of \(\mathrm{GMS}(p)\) over the volume, where \(m_I(p)\) and \(m_{I_{GT}}(p)\) denote the Prewitt operator gradient magnitudes of the images at voxel \(p\), and the stabilizing constant T is set to 170 for dynamic range $L = 255$.

\subsection{Performance and Comparison with Other Methods}
As shown in Fig.~\ref{fig3}, the capillaries in the depth-plane image from single scan (input) exhibit discontinuity and noise, making it difficult to study the intricate retinal vasculature. Without access to long-range dependencies, CNNs struggle to reconstruct capillary segments consistently, particularly in regions with high noise or weak morphological cues, leading to discontinuities of vessels (as the capillary indicated by the arrow in Fig.~\ref{fig3}). In contrast, Transformer-based models leverage self-attention mechanisms to integrate global vascular context, allowing for more accurate inference of fragmented capillaries and better alignment with adjacent structures. . This ability to capture long-range dependencies results in superior capillary reconstruction and vascular continuity in Fig.~\ref{fig3}, consistent with prior work using vision Transformers for vessel segmentation~\cite{xu2024g2vit}. Among all evaluation metrics (Table~\ref{table1}), Freqformer greatly surpassed CNN models and outperformed other transformer-based models, achieving the best quality of capillary reconstruction. Additionally, among the transformer-based models, Freqformer requires the least GPU memory during training, over half reduction to that of Xformer or Restormer, and $\sim$30\% reduction to that of second-best method (ART). This efficient memory usage highlights Freqformer’s advantage in balancing performance and space complexity efficiency.

\begin{table}[!ht]
\caption{Comparison of Different Models in Enhancing Retinal Vasculature}
\label{table1}
    \centering
    \resizebox{\columnwidth}{!}{%
    \begin{tabular}{c c c c c c c c c}
    \specialrule{0.8pt}{0pt}{1pt}
        \hline
        Type & Method & PSNR↑ & SSIM↑ & LPIPS↓ & 3D-SSIM↑ & 3D-GMSD↓ & GMACs (G) & PGMU (GB)  \\ \hline
        \multirow{4}{*}{CNN} & DnCNN\tablefootnote{\url{https://github.com/SaoYan/DnCNN-PyTorch}}\cite{zhang2017beyond} & 23.49 & 0.632 & 0.263 & 0.599 & 0.241 & 88.66 & 1.30  \\ 
        ~ & RIDNet\tablefootnote{\url{https://github.com/saeed-anwar/RIDNet}}\cite{anwar2019real} & 24.64 & 0.679 & 0.221 & 0.629 & 0.199 & 238.74 & 2.32 \\
        ~ & SADNet\tablefootnote{\url{https://github.com/JimmyChame/SADNet}}\cite{chang2020spatial} & 25.22 & 0.697 & 0.212 & 0.645 & 0.193 & 43.08 & 1.01 \\ 
        ~ & NAFNet\tablefootnote{\url{https://github.com/megvii-research/NAFNet}}\cite{chen2022simple} & 26.09 & 0.723 & 0.205 & 0.685 & 0.184 & 38.98 & 3.47 \\ \hline
        \multirow{4}{*}{Transformer} & Restormer\tablefootnote{\url{https://github.com/swz30/Restormer}}\cite{zamir2022restormer} & 25.54 & 0.706 & 0.205 & 0.662 & 0.193 & 250.00 & 20.41  \\
        ~ & Xformer\tablefootnote{\url{https://github.com/gladzhang/Xformer}}\cite{zhang2023xformer} & 26.44 & 0.740 & 0.198 & \underline{0.710} & \underline{0.181} & 325.26 & 20.52  \\
        ~ & ART\tablefootnote{\url{https://github.com/gladzhang/ART}}\cite{zhang2023accurate} & \underline{26.58} & \underline{0.743} & \underline{0.197} & 0.709 & 0.182 & 243.40 & 13.87  \\
        ~ & Freqformer & \textbf{26.90} & \textbf{0.754} & \textbf{0.193} & \textbf{0.741} & \textbf{0.169} & 248.70 & 9.59 \\ \specialrule{0.8pt}{0pt}{0pt}
    \end{tabular}%
    }
\end{table}

\begin{table}[!ht]
\caption{Results of Ablation Study on Freqformer Components}
\label{table2}
    \centering
    \resizebox{\columnwidth}{!}{%
    \begin{tabular}{c c c c c c c c c}
    \specialrule{0.8pt}{0pt}{1pt}
        \hline
        ResBlock & Transformer Layer & CFDM & PSNR↑ & SSIM↑ & LPIPS↓ & 3D-SSIM↑ & 3D-GMSD↓\\ \hline
        \checkmark &   &   & 25.88 & 0.717 & 0.206 & 0.673 & 0.191\\
          & \checkmark &   & 26.19 & 0.728 & 0.203 & 0.703 & 0.184\\
          & \checkmark & \checkmark & 26.90 & 0.754 & 0.193 & 0.741 & 0.169\\ \specialrule{0.8pt}{0pt}{0pt}
    \end{tabular}%
    }
\end{table}

\subsection{Ablation Study and Comparison of Architectural Variants}
To validate the contributions of each component in Freqformer, we conducted ablation experiments on the proposed Transformer layer and the CFDM. To establish a baseline, we replaced all Freqformer blocks with Residual Blocks (ResBlock). As shown in Table~\ref{table2}, the integration of CFDM notably enhances reconstruction performance across all metrics. While the Transformer layer alone also provides meaningful improvements by effectively capturing global spatial interactions, the introduction of CFDM further amplified these gains, highlighting CFDM's substantial and significant role in the network. The synergistic contributions of Transformer layer and CFDM ensure a more balanced and comprehensive feature representation, leveraging the complementary strengths of global spatial and frequency domain information.

\begin{figure}[!t]
\centerline{\includegraphics[width=0.95\columnwidth]{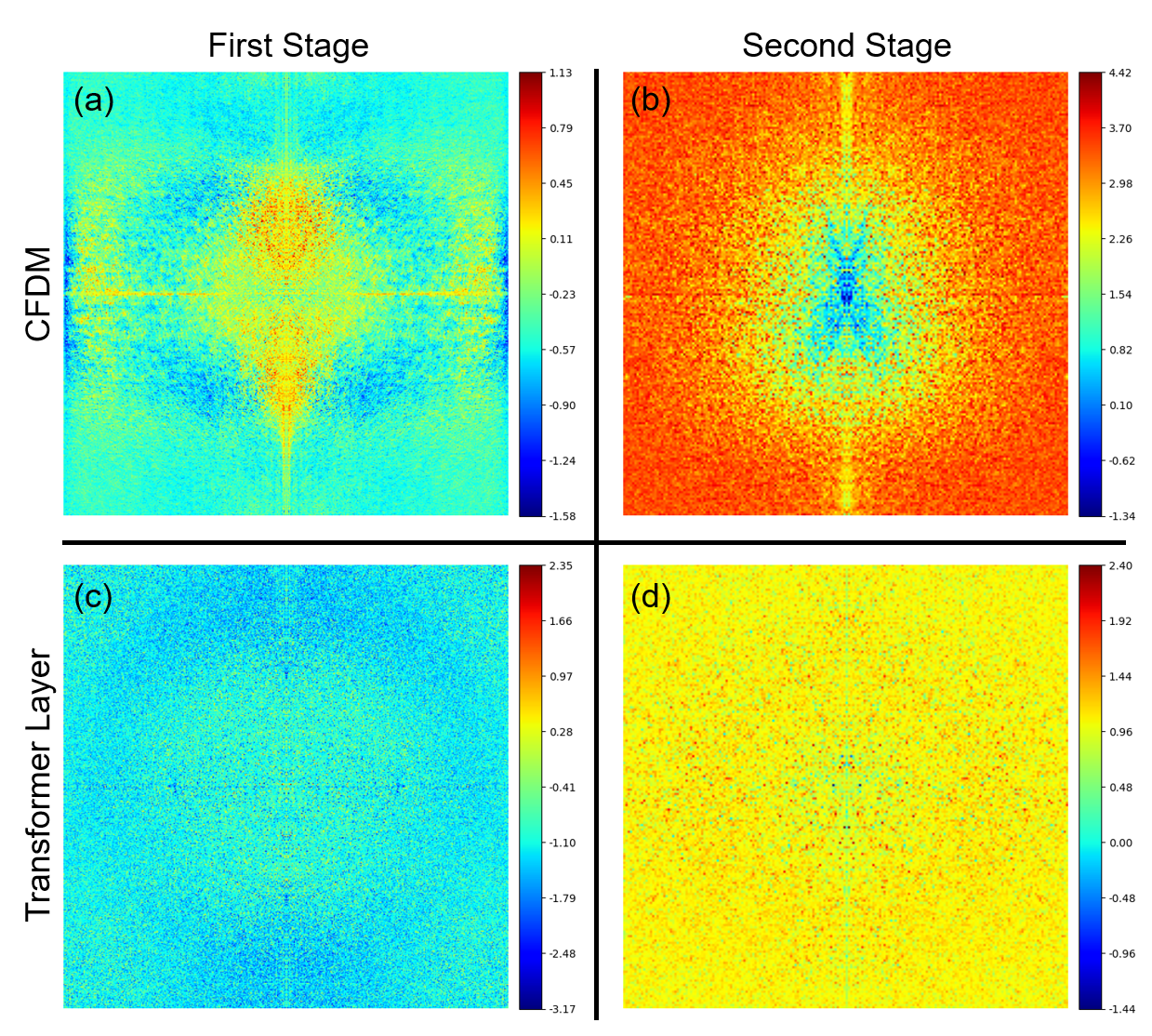}}
\caption{Log-magnitude spectrum difference maps for the CFDM at (a) the first stage and (b) the second stage, as well as for the Transformer layer at (c) first stage and (d) second stage from the first channel of a representative block.}
\label{fig4}
\end{figure}

To further illustrate CFDM's selective frequency-domain enhancement, its functionality is analyzed through log-magnitude spectrum difference maps, which quantify the changes in frequency components induced by CFDM~(Fig.~\ref{fig4}). Specifically, these difference maps are computed by subtracting the log-magnitude spectrum of the feature maps before and after CFDM processing for each channel individually. This approach allows us to visualize how CFDM selectively enhances certain frequency components. In these maps, a strong change (red regions) in specific frequency ranges, such as mid- or high-frequency components, indicates that CFDM actively amplifies those frequency features. As shown in Fig.~\ref{fig4}, CFDM exhibits distinct frequency-dependent enhancement at different network stages, with unique amplification patterns observed at each stage. This confirms that CFDM dynamically adapts its enhancement strategy based on feature representation needs. In contrast, the Transformer layer does not demonstrate this frequency-specific enhancement, as it primarily extracts spatial features. This distinction highlights the complementary roles of the CFDM and Transformer layer—where the Transformer layer preserves global spatial continuity while CFDM refines vascular structures by selectively enhancing critical frequency components.

\begin{figure}[!t]
\centerline{\includegraphics[width=1\columnwidth]{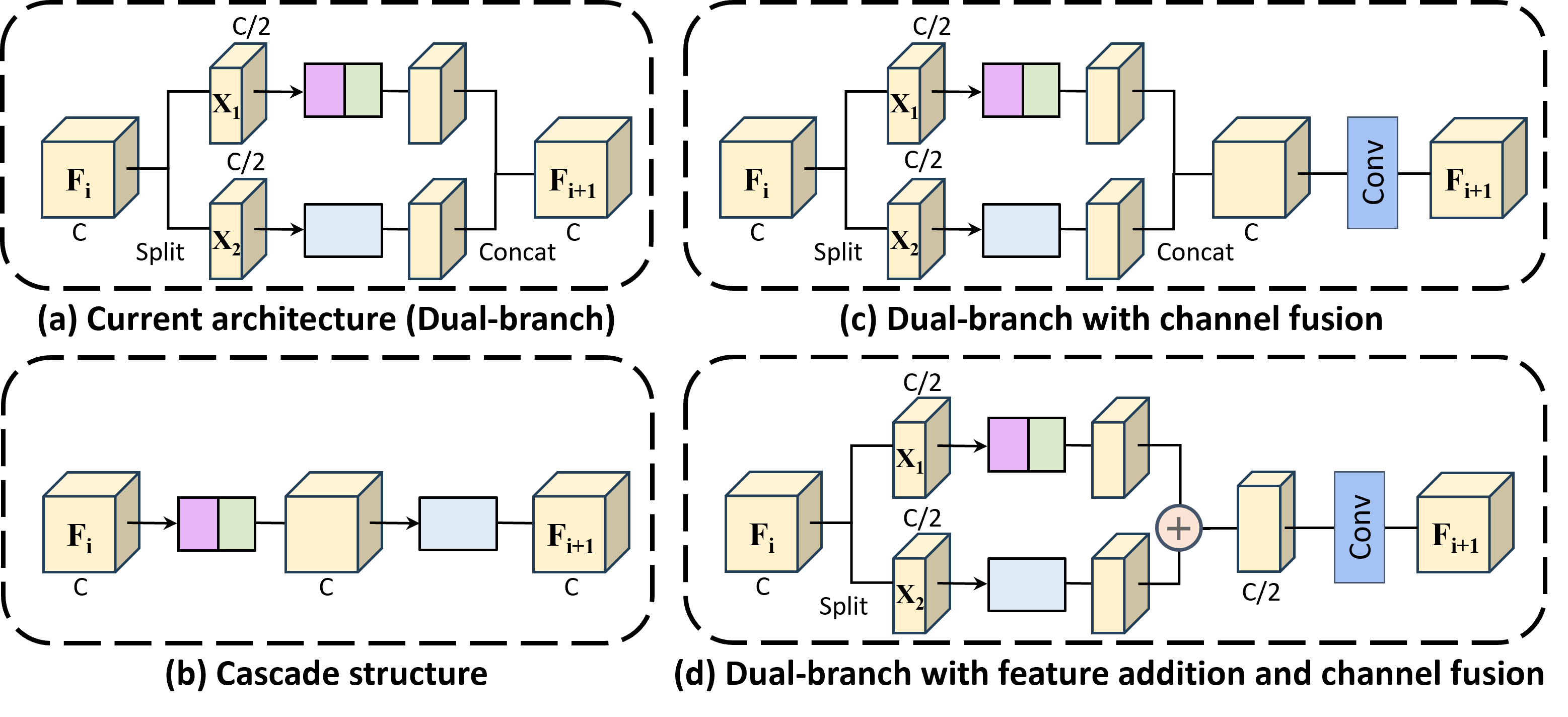}}
\caption{Architectural variants of the Freqformer block.}
\label{fig5}
\end{figure}

Furthermore, to validate the effectiveness of our current dual-branch architecture in leveraging the complementary strengths of the Transformer layer and CFDM, three alternative structures are compared with our design (Fig.~\ref{fig5}). Specifically, model (b) works by cascading the Transformer layer and the CFDM. Model (c) and (d) incorporate two common dual branch feature fusion mechanisms: (c) the channel fusion and (d) feature addition and channel fusion. As presented in Table.~\ref{table3}, the results demonstrate that our design (dual-branch architecture without feature fusion) outperforms other variants across all the metrics. This superiority underscores the critical role of block architecture in effectively integrating spatial and frequency representations, demonstrating that the interference between spatial and frequency features can degrade performance or even negatively impact overall reconstruction quality.

\begin{figure*}[!t]
\centerline{\includegraphics[width=0.9\textwidth]{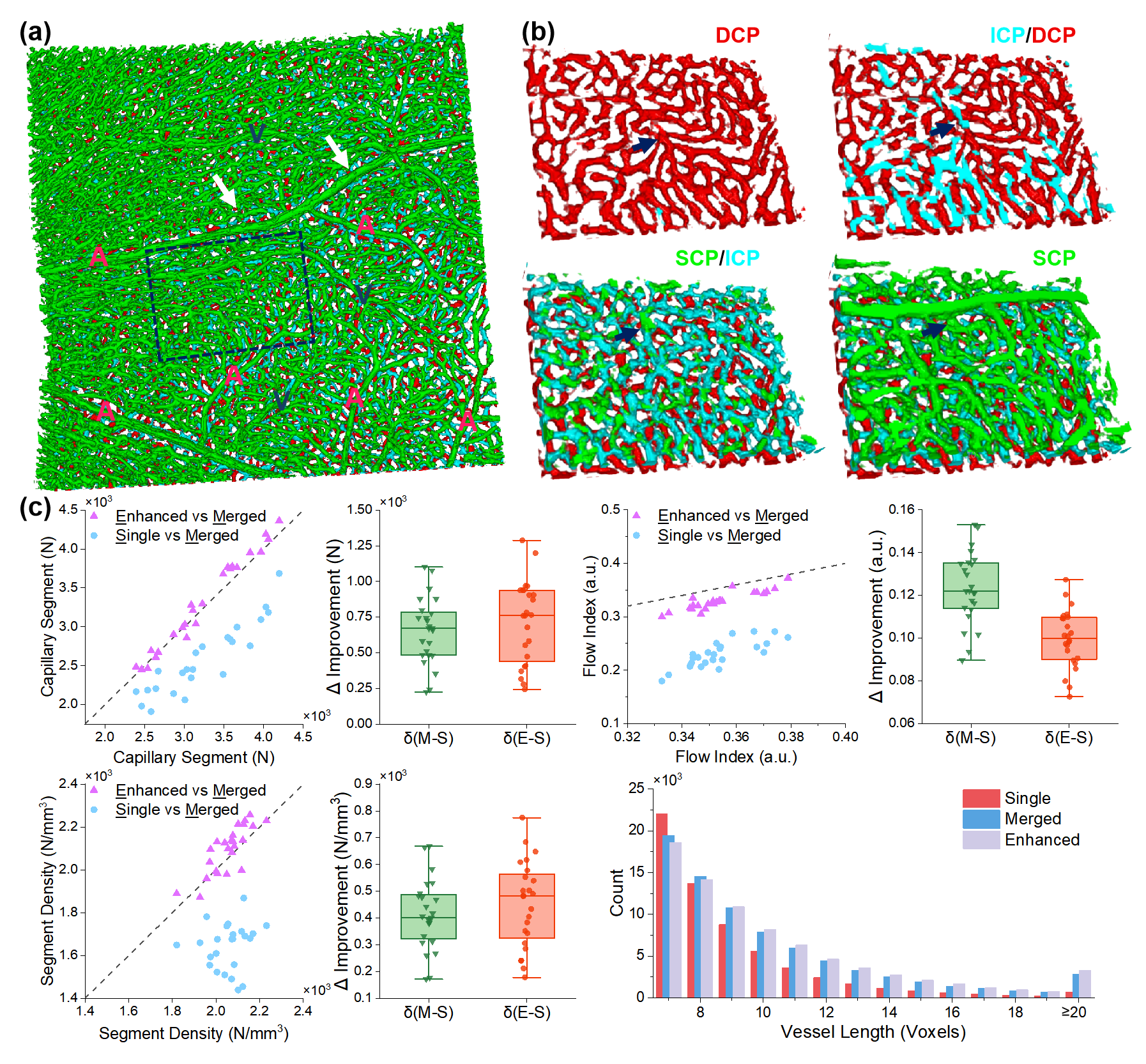}}
\caption{3-D visualization and quantification of retinal vasculature. (a) Retinal vasculature from enhanced OCTA volume, with SCP, ICP, DCP and inter-plexus capillaries color-coded in green, cyan, red, and combination of adjacent colors respectively. Avascular zones (arrows) were observed near arteries (A), but not in veins (V). (b) Zoomed-in view of 3-D retinal vasculature at multiple depth, highlighting a vortex in DCP and its connection to veins in SCP. (c) Capillary segment count, density, flow index, and length histogram calculated in 3-D, as well as the comparisons between single scan, merged scan, and enhanced scan OCTA volumes (N=24).}
\label{fig6}
\end{figure*}

\subsection{3-D Visualization and Quantification of Retinal Microvasculature} 
As shown in Fig.~\ref{fig6}, Freqformer enhanced OCTA volume from single scan achieves detailed 3-D visualization of retinal microvasculature. Specifically, avascular zones are clearly identified adjacent to arteries (indicated by white arrows), but absent near veins (Fig.~\ref{fig6} (a)). Capillary vortices, which are a spider-like arrangement of capillaries in the deep capillary plexus (DCP), are resolved with their connection confidently traced to retinal venules (Fig.~\ref{fig6} (b)). To quantify the retinal microvasculature in 3-D, capillary segment count, segment density, segment length, and flow index within inner retinal volume are calculated and analyzed. Both the enhanced (E) OCTA volumes and merged (M) OCTA volumes resolve more capillary segments (\(\delta\) (M-S)/M: 20\%, and \(\delta\) (E-S)/M: 22\%) than the original single-scan (S) OCTA (Fig.~\ref{fig6} (c)). Statistical analysis showed no significant difference between the enhanced and merged volumes in capillary segment count (p = 0.906) or segment density (p = 0.411), suggesting that Freqformer effectively reconstructs the retinal vasculature to a level comparable with merged volumes. Additionally, consistent with merged volumes, enhanced OCTA volumes demonstrate a higher percentage of longer capillary segments, validating the improved continuity of the vascular network. Compared to the single-scan volumes, the enhanced volumes exhibit an increased flow index (average signal value within 3-D blood vessel masks), approaching ideal levels to the merged volumes (binarized vessel mask obtained from merged volumes and consistent across all three volumes). Furthermore, NRMSE scores between enhanced and merged volumes are calculated for all 3-D vascular quantification metrics to evaluate the quantification error, with Freqformer achieving the lowest NRMSE across all methods (Table~\ref{table4}). Concretely, relative to the strongest Transformer baseline (Xformer), Freqformer reduces NRMSE by $\sim$7\% for segment count, $\sim$10\% for segment density, $\sim$4\% for segment length, and $\sim$33\% for flow index; versus the best CNN (NAFNet), the reductions are $\sim$32\%, $\sim$37\%, $\sim$31\%, and $\sim$35\%, respectively. Notably, Freqformer’s largest margin is on the flow index, indicating that the benefits extend beyond morphological continuity to a more faithful recovery of flow-related OCTA signal intensities. Taken together, these results have direct practical implications. Achieving near-merged-volume fidelity from a single acquisition removes the need for multi-volume merging—rarely feasible in routine clinics, especially for large-FOV protocols—thereby shortening examination time, reducing motion/registration failures, and lowering patient burden (children, elderly, poor fixation). Moreover, the improved capillary continuity and recovered OCTA signal stabilize downstream 3-D vascular quantification, which improves repeatability; in longitudinal or treatment-response settings this translates into a smaller minimal detectable change and greater statistical power for the same cohort size, enabling finer tracking of disease trajectories and therapy effects.

\begin{table}[!ht]
\caption{Comparisons of different architectural variants}
\label{table3}
    \centering
    \resizebox{\columnwidth}{!}{%
    \begin{tabular}{c c c c c c}
    \specialrule{0.8pt}{0pt}{1pt}
        \hline
        Architecture & PSNR↑ & SSIM↑ & LPIPS↓ & 3D-SSIM↑ & 3D-GMSD↓ \\ \hline
        Model (a) & 26.90 & 0.754 & 0.193 & 0.741 & 0.169 \\
        Model (b) & 25.43 & 0.697 & 0.207 & 0.678 & 0.187 \\
        Model (c) & 24.69 & 0.671 & 0.220 & 0.613 & 0.201 \\
        Model (d) & 26.08 & 0.715 & 0.206 & 0.678 & 0.189\\ \specialrule{0.8pt}{0pt}{0pt}
    \end{tabular}%
    }
\end{table}

\begin{table}[!ht]
\caption{Comparison of all models on the normalized root mean squared error (NRMSE) scores calculated between the enhanced volume and ground truth for each 3-D vascular quantification metric}
\label{table4}
    \centering
    \resizebox{\columnwidth}{!}{%
    \begin{tabular}{c c c c c c c c c c}
    \specialrule{0.8pt}{0pt}{1pt}
        \hline
        Quantification & Freqformer & Single & DnCNN & RIDNet & SADNet & NAFNet & Restormer & Xformer & ART \\ \hline
        Capillary Segment & \textbf{0.0367} & 0.2138 & 0.0892 & 0.0608 & 0.0580 & 0.0539 & 0.0504 & \underline{0.0393} & 0.0413 \\
        Segment Density & \textbf{0.0354} & 0.2105 & 0.0971 & 0.0589 & 0.0565 & 0.0584 & 0.0473 & \underline{0.0393} & 0.0404 \\ 
        Length & \textbf{0.0507} & 0.3906 & 0.3769 & 0.0989 & 0.0732 & 0.0705 & 0.0699 & \underline{0.0530} & 0.0634 \\ 
        Flow Index & \textbf{0.0710} & 0.3519 & 0.1974 & 0.1971 & 0.1730 & 0.1085 & 0.1342 & \underline{0.1062} & 0.1078 \\ \specialrule{0.8pt}{0pt}{0pt}
    \end{tabular}%
    }
\end{table}

\section{Discussion}
\subsection{Advantages of Enhanced 3-D OCTA Volume}
Traditionally, OCTA scans are analyzed using projected 2-D \textit{en face} images over a specified slab (depth range). These \textit{en face} images offer higher SNR than single depth-plane images, making them well-suited for visualizing retinal microvasculature and training deep learning models for enhancement~\cite{gao2020reconstruction, kadomoto2020enhanced}. Unfortunately, the depth information is collapsed in these 2-D images, thereby hindering the appreciation and quantification of 3-D retinal vascular network. Although some studies have explored 3-D visualization of OCTA, they have primarily focused on segmenting binary masks of blood vessels~\cite{sampson2022towards}. This approach neglects the de-correlation angiogram information inherent to OCTA volume, which is closely linked to blood flow rate~\cite{su2016calibration} and serves as a robust biomarker for retinal diseases such as diabetic retinopathy (DR)~\cite{dadzie2023normalized}.

In contrast, the enhanced high-definition OCTA volumes generated by Freqformer preserve flow information for comprehensive volumetric analysis of the retinal microvasculature. Here, with Freqformer, we revealed the organization of retinal arteries and veins and extracted capillary segments in 3-D (Fig.~\ref{fig6}). Using these data provided with commercial OCT, we successfully quantified the capillary segment density \textit{in vivo} in 3-D for human retinas for the first time. In the future, we hope that more detailed quantification of 3-D microvasculature would provide us more insights into the blood flow and retinal tissue oxygenation\cite{corliss2019methods}.

\subsection{2-D instead of 3-D Patches for Reconstruction}
In this study, we opted for 2-D single-depth-plane angiogram images rather than 3-D patches for training and reconstruction due to both computational constraints and feature extraction limitations. First, 3-D patch approaches introduce substantial computational overhead, making them impractical for real-time clinical applications. OCTA volumes are inherently high resolution, and applying 3-D convolutions or MHA across patches would exponentially increase GPU memory requirements and inference time, making it infeasible for clinical deployment. Second, while 3-D convolutions can theoretically enhance spatial continuity, they assume uniform information density across all dimensions. However, in OCTA, the en-face (lateral) images contain much richer vascular contrast than cross-sectional B-scans (axial direction). If 3-D patches were used, high-contrast vascular features from en-face images could be diluted by low-contrast axial features, making it inefficient for extracting capillary details \cite{huang2024upping}. 

To assess the relative effectiveness of 2-D versus 3-D networks, we implemented 3-D counterparts of Freqformer as well as other networks in Table~\ref{table1}. To be specific, all 2-D convolutional layers, pixel shuffle, layer normalization, pooling operations and custom modules (e.g., FFT or masking strategy) were replaced with their 3-D counterparts while preserving the original configurations. In addition, MHA and MLP in Transformer layers were adapted to process volumetric tokens without altering the original formulation. Following common practice for 3-D networks, we adopted a patch-based training and inference scheme to accommodate GPU memory limits. We used the same 20 volumetric scans as in the 2-D experiments for training and validation. We first localized the retinal depth range in each volume using our OCT B-scan segmentation algorithm, and then extracted patches exclusively from this retinal region to avoid excessive background in OCTA. Each OCTA retinal volume was partitioned into non-overlapping 3-D patches of size 32\(\times\)64\(\times\)64 (Depth\(\times\)Height\(\times\)Width) and split 9:1 for training and validation. For testing, we used the same 24 scans as in the 2-D experiments. During inference, we performed dense sliding-window evaluation with 32\(\times\)64\(\times\)64 patches and strides of 16\(\times\)28\(\times\)28, where overlapping voxel predictions were averaged to mitigate boundary artifacts and ensure spatial consistency. 

\begin{figure}[!t]
\centerline{\includegraphics[width=1\columnwidth]{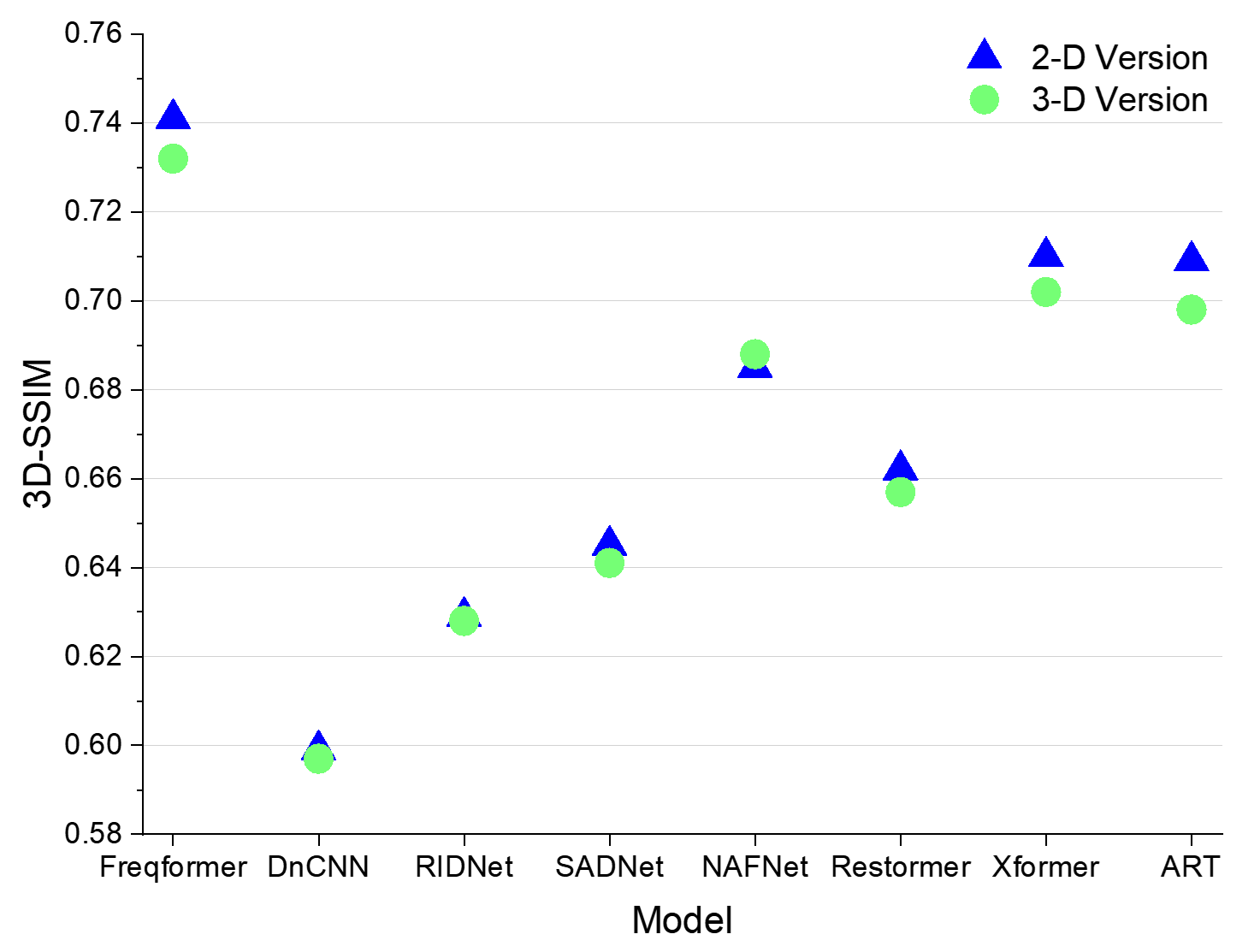}}
\caption{Comparisons between 2-D and 3-D networks on 3D-SSIM. The blue triangle symbols represent the original 2-D networks, while the green circles denote their 3-D counterparts.}
\label{fig7}
\end{figure}

As shown in Fig.~\ref{fig7}, converting the networks to 3-D does not change the coarse performance hierarchy and Freqformer achieves best performance in both 2-D and 3-D, indicating that its advantage is not tied to the dimensionality of the backbone. The extent of the performance change associated with the transition from 2-D to 3-D varies across model types. For CNN baselines, the change is modest, with only small deviations between the 2-D and 3-D versions. In contrast, Transformer-based models exhibit more pronounced variation when moved to 3-D. This is likely because flattening 3-D patches into token sequences in MHA breaks the native volumetric geometry, making it harder for Transformer-based networks to model geometric relationships even with relative positional embeddings. In addition, we selected the top CNN baseline (NAFNet) and the strongest Transformer baseline (Xformer) excluding Freqformer, together with their 3-D counterparts, for head-to-head comparisons on inference time, image quality and NRMSE of 3-D vascular quantification. As shown in Table~\ref{table5}, the 3-D variants do not improve image metrics or quantification error and instead incur nearly 10\(\times\) longer inference time than their original version, which is partly attributable to overlapping sliding-window strategy. These results support our design choice to enhance volumes slice by slice with Freqformer, which offers the best balance of image quality, 3-D vascular quantification accuracy, runtime, and GPU requirement.

\begin{table}[!ht]
\caption{Comparison of 2-D and 3-D network performance on metrics and NRMSE of 3-D vascular quantification}
\label{table5}
    \centering
    \resizebox{\columnwidth}{!}{%
    \begin{tabular}{c c c c c c c}
    \specialrule{0.8pt}{0pt}{1pt}
        \hline
        Metrics & Freqformer & Xformer & NAFNet & 3D-Freqformer & 3D-Xformer  & 3D-NAFNet \\ \hline
        Inference Time per Volume (s) & 44.9 & 68.9 & 15.1 & 581.7 & 641.2 & 229.9 \\
        3D-SSIM↑ & 0.741 & 0.710 & 0.685 & 0.732 & 0.702 & 0.688 \\
        3D-GMSD↓ & 0.169 & 0.181 & 0.184 & 0.172 & 0.182 & 0.184 \\
        Capillary Segment (NRMSE) & 0.0367 & 0.0393 & 0.0539 & 0.0401 & 0.0442 & 0.0526 \\
        Segment Density (NRMSE) & 0.0354 & 0.0393 & 0.0584 & 0.0413 & 0.0451 & 0.0568 \\
        Length (NRMSE) & 0.0507 & 0.0530 & 0.0705 & 0.0544 & 0.0601 & 0.0701 \\
        Flow Index (NRMSE) & 0.0710 & 0.1062 & 0.1085 & 0.0813 & 0.1076 & 0.1080\\ \specialrule{0.8pt}{0pt}{0pt}
    \end{tabular}%
    }
\end{table}

\begin{figure*}[!t]
\centerline{\includegraphics[width=0.9\textwidth]{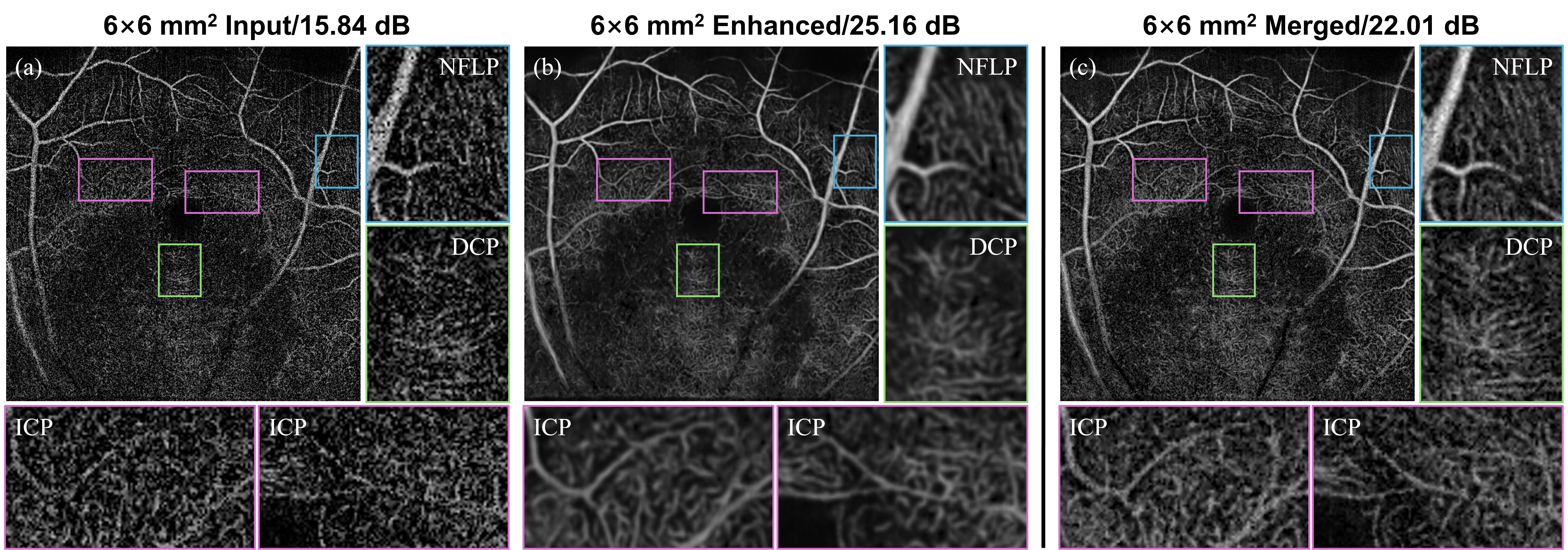}}
\caption{Generalization results of Freqformer (trained with 3\(\times\)3 \(\text m \text m^{\text 2}\) scans) on the 6\(\times\)6 \(\text m \text m^{\text 2}\) scan with SNR listed. (a) Single-scan image. (b) Enhanced image. (c) Merged image (N=10). Four regions of interest are zoomed in, covering nerve fiber layer plexus (NFLP), intermediate capillary plexus (ICP), and deep capillary plexus (DCP). }
\label{fig8}
\end{figure*}

\subsection{Generalization Capability}
To further evaluate the generalization performance of Freqformer, we applied it to OCTA images acquired using the same device but with an expanded field of view (FOV, 6\(\times\)6 \(\text m \text m^{\text 2}\)). Large FOV scans are increasingly used in clinical settings to capture a broader range of pathological features, such as non-perfusion areas, microaneurysms, and neovascularization ~\cite{zhang2016wide}. However, with a fixed A-line scanning rate and the same or slightly increased A-line number, expanding the FOV results in lower resolution, leading to decreased image quality of retinal microvasculature. While volume merging can partially compensate for the reduced quality, the improvement is often limited, and precise registration is further complicated by scan-to-scan displacements due to fixation shifts. As shown in Fig.~\ref{fig8}, merging 10 volumes from 6\(\times\)6 \(\text m \text m^{\text 2}\) scans partially improves vascular connectivity and signal-to-noise ratio (SNR)~\cite{wang2022compressive}, defined as the ratio of the mean signal within vascular regions to the standard deviation of noise in the foveal avascular zone, compared to a single scan.  However, the resulting image still shows a lower SNR than the standard 3\(\times\)3 \(\text m \text m^{\text 2}\) merged image. In contrast, Freqformer reconstructs capillary structures with high morphological consistency to the merged reference while also achieving higher SNR values (\(\Delta\)=1.98 ± 0.59 (SD), p$<$0.001, N=10). These findings demonstrate Freqformer’s capability to perform super resolution imaging for large FOV scans (such as 12\(\times\)12 \(\text m \text m^{\text 2}\)), offering a viable alternative to conventional volume-merging workflows. Its robustness and versatility make it a promising tool for broader clinical applications.

\subsection{Limitations}
While the results demonstrate substantial improvements of Freqformer in both visual and quantitative performance compared to other methods, there are several limitations that may affect its broader applicability and impact. First, Our study used PLEX® Elite 9000 SS-OCT for OCTA acquisition. While we demonstrated generalization across different scanning densities and fields of view, the model’s performance on other OCTA devices (e.g., SD-OCT or different commercial systems) remains unverified. Further validation across diverse imaging modalities and hardware platforms is needed. Second, this study primarily focused on healthy subjects, which may limit the model’s generalizability to pathological cases with vascular abnormalities such as capillary dropout, neovascularization, and vessel tortuosity. Evaluating Freqformer on diseased retinas will be necessary to assess its robustness in real-world clinical applications. Third, while Freqformer effectively reconstructs depth-resolved vasculature, direct quantitative validation of full 3-D vascular topology remains challenging due to the lack of a well-established ground truth for volumetric vascular structures in OCTA. Future work should integrate automated vessel segmentation and topology analysis to improve 3-D vasculature validation. Finally, despite our efforts to optimize space complexity, the computational demand of Freqformer is still higher than that of CNNs. Further research into streamlining the architecture for less powerful GPU would improve Freqformer's accessibility and ease of use.

\section{Conclusion}
In conclusion, this study presents Freqformer, a novel frequency-domain Transformer architecture designed to enhance OCTA volumes and thus the entire retinal vasculature in 3-D. By integrating complex-valued frequency-domain module and cross-covariance attention mechanism, Freqformer effectively addresses limitations of conventional OCTA processing, delivering detailed, high-definition visualization of retinal microvasculature with improved capillary continuity and reduced noise. Experimental results demonstrate that Freqformer outperforms state-of-the-art CNN-based and Transformer-based methods, offering superior enhancement and enabling capillary segment-wise quantification in human retinas. Additionally, Freqformer displays great generalizability on the 6\(\times\)6 \(\text m \text m^{\text 2}\) protocol. Despite limitations, like the lack of training on pathological data and requirement of high-definition ground truth during training, Freqformer shows significant promise as a powerful tool for retinal imaging. Future efforts will focus on optimizing Freqformer for clinical deployment, enhancing its efficiency, and expanding its capabilities for flow quantification and analysis of diseased retinas. These advancements will further solidify Freqformer’s role in advancing the diagnosis and monitoring of retinal vascular diseases.

\bibliographystyle{IEEEtran}
\bibliography{reference}

\end{document}